\documentclass[12pt]{iopart}
\usepackage{latexsym}
\usepackage{graphicx}
\usepackage{amsmath}
\usepackage{amssymb}
\usepackage{amsthm}
\usepackage{amsfonts}
\usepackage{physics}
\usepackage{bm}
\usepackage{dsfont}
\usepackage{caption}
\usepackage{subcaption}
\usepackage{tikz}
\usepackage{hyperref}

\usepackage{csquotes}

\usepackage{indentfirst}
\usepackage{mparhack}
\usepackage{relsize}
\usepackage{lettrine}
\usepackage[capitalise]{cleveref}
\usepackage[%
    hyperref=true,%
    url=false,%
    isbn=false,%
    backref=false,%
    maxcitenames=20,%
    maxbibnames=3,%
    block=none,%
    giveninits=true,%
    sorting=none,%
    backend=bibtex,
]{biblatex}
\begin{document}

\title[Dual unitary circuits in random geometries]{Dual unitary circuits in random geometries}

\author{Yusuf Kasim \& Tomaž Prosen}

\address{Department of Physics, Faculty of Mathematics and Physics, University of Ljubljana, SI-1000 Ljubljana, Slovenia}
\ead{yusuf.kasim@fmf.uni-lj.si tomaz.prosen@fmf.uni-lj.si}
\vspace{10pt}
\begin{indented}
\item[]June 2022
\end{indented}

\begin{abstract}
Recently introduced dual unitary brickwork circuits have been recognised as paradigmatic exactly solvable quantum chaotic many-body systems with tunable degree of ergodicity and mixing. Here we show that regularity of the circuit lattice is not crucial for exact solvability. We consider a circuit where random 2-qubit dual unitary gates sit at intersections of random arrangements of straight lines in two dimensions ({\em mikado}) and analytically compute the variance of the spatio-temporal correlation function of local operators. Note that the average
correlator vanishes due to local Haar randomness of the gates.
The result can be physically motivated for two random mikado settings. The first corresponds to the thermal state of free particles carrying internal qubit degrees of freedom
which experience interaction at kinematic crossings, 
while the second represents rotationally symmetric (random euclidean) space-time.
\end{abstract}


\section{Introduction}
Taming real time dynamics of interacting many-body systems represents one of the main challenges of theoretical physics.
The cleanest and most versatile setting to study interacting many-body dynamics is arguably that of quantum circuits with up to 2-qubit gates (representing local interactions). Although the classical simulation of generic quantum circuits has been in general recognised as intractable, and is currently employed to demonstrate the so-called supremacy of quantum computers~\cite{Google2019}, 
we have recently identified an interesting type of so-called dual unitary circuits where the problem of computing certain dynamical quantities, such as two-point spatio-temporal correlation functions~\cite{Bertini_et_al_PRL_2019,Claeys1,Claeys2,Arul1,Arul2,Prosen21},
quantum quenches~\cite{Piroli},
spectral form factors~\cite{BKP18,CMP}, operator entanglement~\cite{Bertini_et_al_SCIPOST_2020}, etc, can be solved efficiently or even analytically.
Dual unitary circuits, with the gates arranged on a regular brickwork pattern, possess a unique defining feature: specifically, the many-body propagator is not only unitary in the time direction, but can also be identified with a unitary dynamics in the space direction.
Furthermore, one can argue that chaotic variants of dual unitary circuits, with
provable random matrix level statistics and exponentially decaying temporal correlations, represent an exactly solvable instance of ergodic and mixing quantum many-body dynamics. As such, unlike integrable systems, they may be expected to be structurally stable against small perturbations breaking dual-unitarity~\cite{Kos_et_al_PRX_2021}.

Dual unitary circuits studied so far were all defined on regular lattices, either with 2-on-2 gates arranged as vertices of a square lattice (brickwork), or with 3-on-3 gates arranged as vertices of a hexagonal lattice~\cite{Khemani}. 
By factorising these so-called tri-unitary 3-on-3 gate in terms of triples of dual unitary 2-on-2 gates, the latter can in fact be represented as the circuit on the Kagome lattice where 2-qubit dual unitary gates are placed on all intersections of pairs of lines.

In this paper, we propose an additional conceptual modification towards a definition of more general dual unitary systems in 1+1 dimensions. Instead of insisting on a regular pattern of 2-on-2 dual unitary gates, we consider an arbitrary arrangement of straight lines in 1+1 dimensions (mikado). Stipulating that each line (wire) carries a qubit state, we place a (possibly different) dual unitary gate $U$, with $4 \times 4$ matrix $U_{ij}^{kl}$ ($i,j,k,l\in\{1,2\}$) at any intersection.
Requirement of dual-unitarity of $U$, means that both $U_{ij}^{kl}$ and 
$\tilde{U}_{ij}^{kl} = U_{ik}^{jl}$ should be unitary. Assuming that none of the lines (wires) is precisely parallel to the time axis $t={\rm const}$, we thus define a circuit dynamics between times $t=0$ and $t=T$, where the size of the system $N$ -- the number of qubits --- is equal to the total number of lines (Figure~\ref{fig:R_Lat}a).
By requirement of dual-unitarity, such circuit is unitary, irrespective of the orientation of each gate, i.e. which pair of adjacent lines are considered as input/output qubits. 
Requiring gates' unitarity alone, the circuit generally possesses a unique (fixed) direction of time and reduces to the so-called hybrid semiclassical model of Ref.~\cite{Marton}, where qubit/spin degrees of freedom pairwise interact when classical free-particle trajectories cross.

\section{Mikado dual unitary circuits}
\label{Sec:Mikado_dual_unit}
\begin{figure}
\centering
\begin{subfigure}{.5\textwidth}
  \centering
  \includegraphics[width=.9\linewidth]{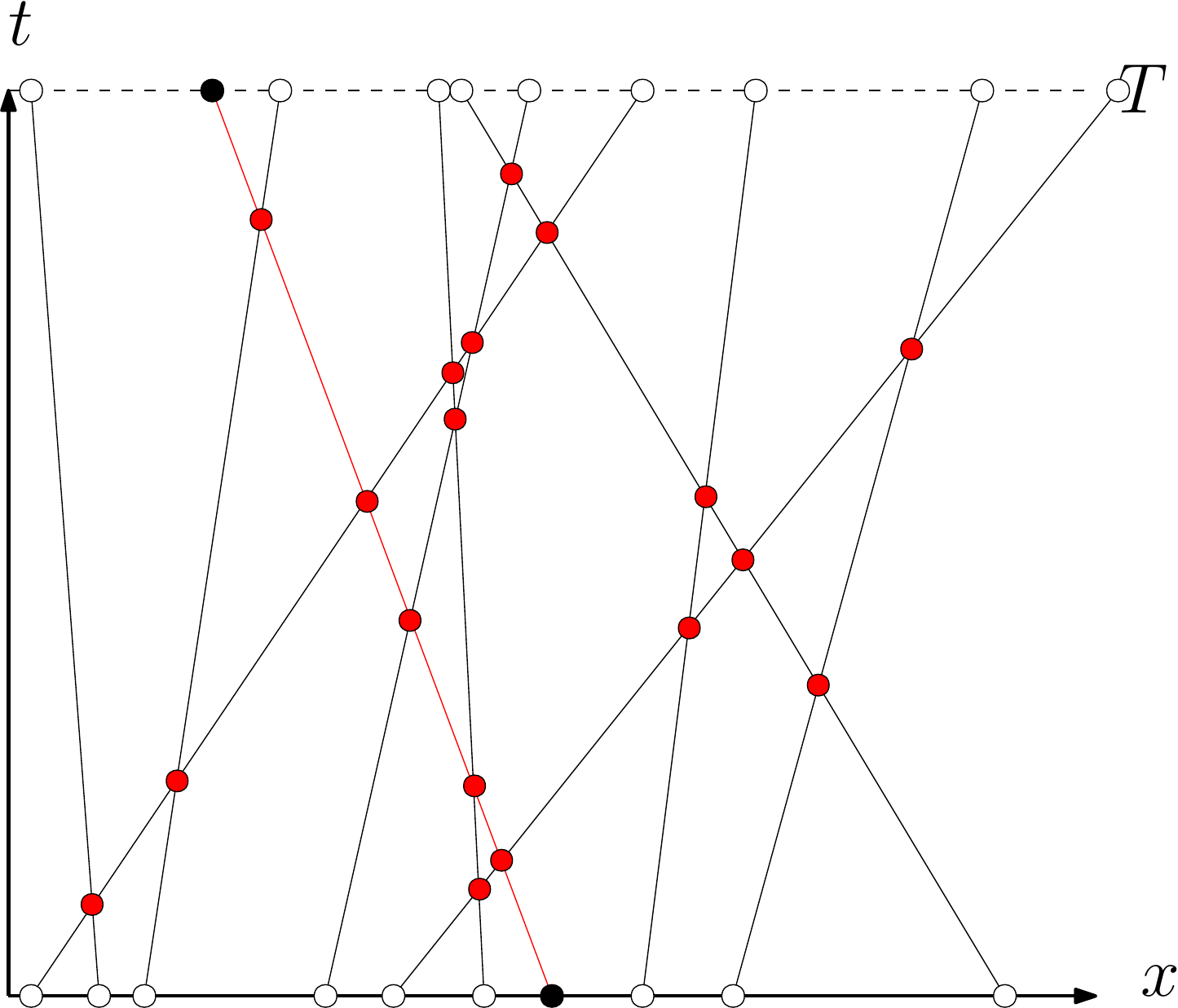}
  \caption{Before contraction}
  \label{fig:R_LatNC}
\end{subfigure}%
\begin{subfigure}{.5\textwidth}
  \centering
  \includegraphics[width=.9\linewidth]{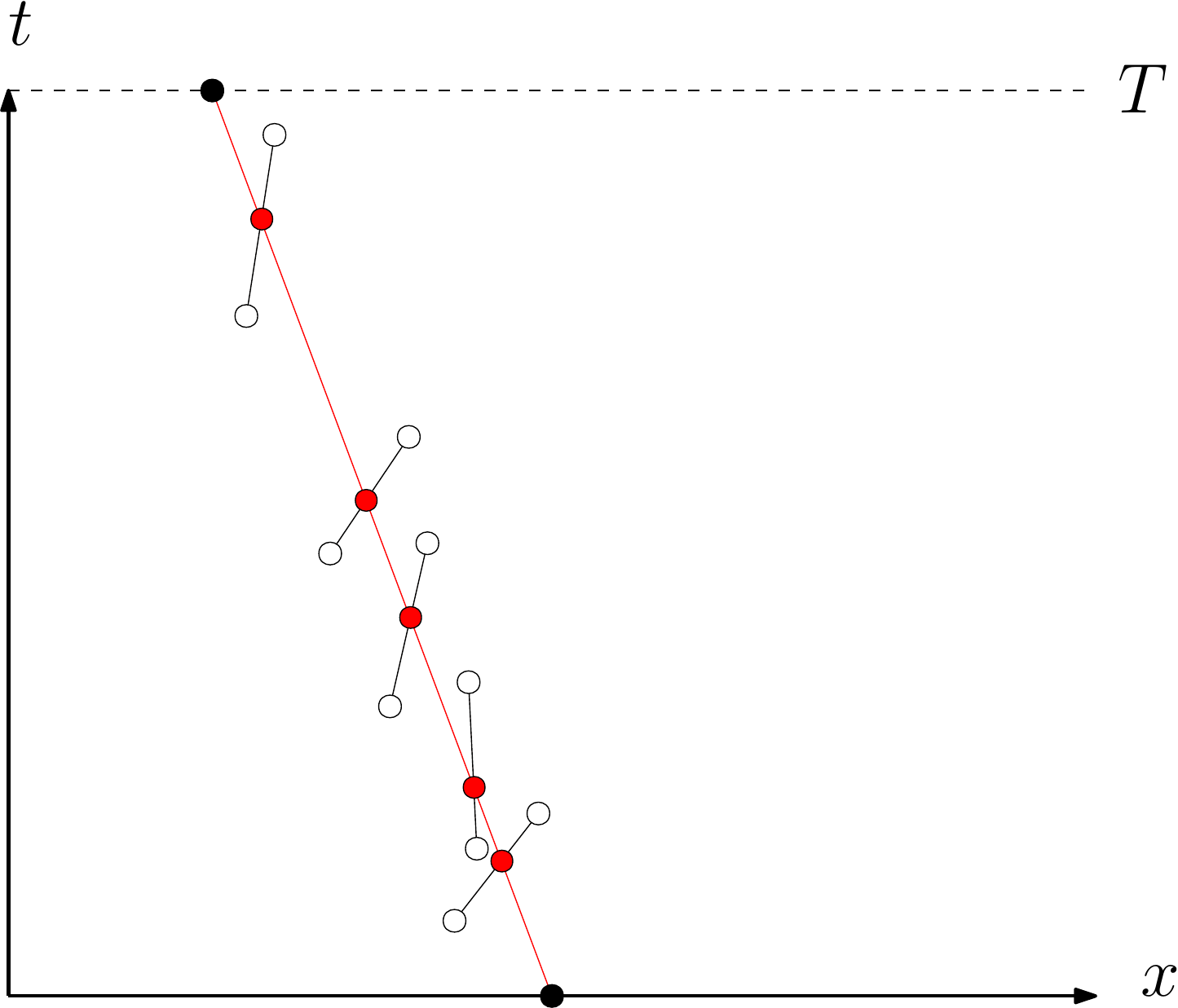}
  \caption{After contraction}
  \label{fig:R_Lat_Cont}
\end{subfigure}
\caption{Example of a local correlation function of a dual unitary random mikado lattice (a). The crossings between wires (red circles) represent dual-unitary gates, while the full circles (bullet) at the initial and final time represent the initial and final local operator. The panel on the right (b) shows the identical tensor network obtained after implementing all dual unitarity conditions (contractions). The correlator is nonzero only if initial and final operator are connected by a straight wire (red line).}
\label{fig:R_Lat}
\end{figure}

 We investigate the behavior of spatio-temporal correlation functions and their decay using random dual unitary gates in random geometries, with an example shown in \cref{fig:R_LatNC}. 
 Let us write the full unitary circuit between times $t_1$ and $t_2$ for a fixed mikado arrangement as $\mathbb U (t_1,t_2) \in {\rm End}((\mathbb C^2)^{\otimes N})$.
 Denoting the $n$-th line coordinate at time $t$ as $x_n(t)$, we define the $k$-th moment of spatiotemporal correlator between a pair of local operators
$a,b\in{\rm End}(\mathbb C^2)$, with local embeddings $a_n,b_n \in {\rm End}((\mathbb C^2)^{\otimes N})$, as
\begin{eqnarray}
C^{(k)}_{ab}(x_{\rm i},t_{\rm i};x_{\rm f},t_{\rm f}) &=& 
\sum_{n,n'}
\delta(x_{\rm i}-x_n(t_{\rm i}))\delta(x_{\rm f}-x_{n'}(t_{\rm f})) \nonumber \\
&&\qquad \times \left\{2^{-N} {\rm tr}
\left(a_{n}\mathbb U(t_1,t_2)^\dagger b_{n'}\, \mathbb U(t_1,t_2)\right)\right\}^k.
\label{eq:cf}
\end{eqnarray}
In the next step we use the folded picture~\cite{MCBanuls_et_al_2009} (following the notation 
of~\cite{Kos_et_al_PRX_2021,Prosen21}), to
rewrite (\ref{eq:cf}) as
\begin{equation}
C^{(k)}_{ab}(x_1,t_1;x_2,t_2) = 
\sum_{n,n'}
\delta(x_1-x_n(t_1))\delta(x_2-x_{n'}(t_2)) \bra{a_n}\mathbb W(t_1,t_2)
\ket{b_{n'}}^k.
\label{eq:cff}
\end{equation}
Here $\mathbb W(t_1,t_2)\in{\rm End}((\mathbb C^4)^{\otimes N})$ is exactly the same unitary circuit as $\mathbb U(t_1,t_2)$ but generated by 
the folded 2-qudit (operator) gates with $d=4$, 
for which we will use a diagrammatic 
representation
\begin{equation}
    W_{(ii')(jj')}^{(kk')(ll')} =
    U_{ij}^{kl}\bar{U}_{i'j'}^{k'l'} =
\begin{tikzpicture}[baseline={([yshift=-.5ex]current bounding box.center)}]
    \node[draw,shape=circle,fill=red,minimum size=20pt] (U) at (0,0) {};
    \node (n1) at (-1,-1) {$ii'$};
    \node (n2) at (-1,1) {$kk'$};
    \node (n3) at (1,-1) {$jj'$};
    \node (n4) at (1,1) {$ll'$};
    \draw[-] (U) -- (n1);
    \draw[-] (U) -- (n2);
    \draw[-] (U) -- (n3);
    \draw[-] (U) -- (n4);
    \end{tikzpicture}
    \label{W}
\end{equation}
The wires thus carry local operator-states, which we will graphically designate by $\bullet$. The self-contraction of the wire (summation over unprimed/bra and primed/ket indices in Eq. (\ref{W})) either represents a unit operator or taking the trace and shall be designated by 
$\circ$.

The property of dual unitarity can then be neatly expressed as a single diagram, stating that any pair of self-contracting wires can be simply pulled thru the interaction vertex (note that a pair of `operator wires' can have a pair of contraction symbols $\circ$ on any side):
\begin{equation}
    \label{eq:contraction}
    \begin{tikzpicture}[baseline={([yshift=-.5ex]current bounding box.center)}]
    \node[draw,shape=circle,fill=red,minimum size=20pt] (U) at (0,0) {};
    \node[draw,shape=circle,fill=white,scale = 0.5] (n1) at (-0.55,-1) {};
    \node (n2) at (-1.4,1) {};
    \node[draw,shape=circle,fill=white,scale = 0.5] (n3) at (1.4,-1) {};
    \node (n4) at (0.55,1) {};
    \draw[-] (U) -- (n1);
    \draw[-] (U) -- (n2);
    \draw[-] (U) -- (n3);
    \draw[-] (U) -- (n4);
    \end{tikzpicture}
     = 
    \begin{tikzpicture}[baseline={([yshift=-.5ex]current bounding box.center)}]
    \node[draw,shape=circle,fill=white,scale = 0.5] (n1) at (0.5,0) {};
    \node (n2) at (-2.1,1) {};
    \node[draw,shape=circle,fill=white,scale = 0.5] (n3) at (-0.5,0) {};
    \node (n4) at (1.15,1) {};
    \draw[-] (n1) -- (n4);
    \draw[-] (n3) -- (n2);
    \end{tikzpicture}
\end{equation}
Dual unitary gates of qubits can be completely parametrised as \cite{Bertini_et_al_PRL_2019}:
\begin{equation}
    \label{du_unit}
    \begin{split}
        & U = e^{i\phi} (u_+ \otimes u_-) V[J] (v_- \otimes v_+) \\
        & V[J] = \exp \left[ -i \left( \frac{\pi}{4} \sigma^x \otimes \sigma^x + \frac{\pi}{4} \sigma^y \otimes \sigma^y + J \sigma^z \otimes \sigma^z \right) \right],
    \end{split}
\end{equation}
where $\sigma^{x,y,z}$ are the Pauli matrices, 
$u_\pm , v_\pm \in \mathrm{SU}(2)$ are arbitrary local 1-qubit gates, and $J \in [0,\frac{\pi}{4}]$ is an interaction parameter which uniquely determines the entangling power of the gate~\cite{Arul1}. For instance, $J=\pi/4$ corresponds to the non-interacting (SWAP) gate, while $J=0$ yields the case of the {\em maximal chaos}~\cite{Bertini_et_al_SCIPOST_2020}.

Diagram \cref{eq:contraction} can be used to drastically simplify the circuit of 
\cref{fig:R_LatNC}. Assuming that the operators
$a,b$ are traceless, the diagram of \cref{fig:R_LatNC} contracts to zero, unless we place the operators $a,b$ on the same line/wire. In the latter case, the correlator reduces to the product of unital 1-qubit quantum channels, as depicted in \cref{fig:R_Lat_Cont}.
These local unital channels, written in the 3-dimensional Pauli basis $\{\sigma^\alpha;\alpha=x,y,z\}$ -- while on the unit operator $\mathds{1}$ they act trivially -- are represented with elementary diagrams
\begin{equation}
\label{MMatrix}
    \mathcal{M}[u,v] = 
     R[u]\,D\,R[v]
     =
    \begin{tikzpicture}[baseline={([yshift=-.5ex]current bounding box.center)}]
    \node[draw,shape=circle,fill=red,minimum size=15pt] (U) at (0,0) {};
    \node[draw,shape=circle,fill=white,scale = 0.5] (n1) at (-0.55,-1) {};
    \node (n2) at (-1.4,1) {};
    \node (n3) at (1.4,-1) {};
    \node[draw,shape=circle,fill=white,scale = 0.5] (n4) at (0.55,1) {};
    \node[draw,shape=circle,fill=cyan] (u) at (0.7,-0.5) {$v$};
    \node[draw,shape=circle,fill=cyan] (v) at (-0.7,0.5) {$u$};
    \node (arin1) at (1.4,-1) {};
    \node (arin2) at (0.91,-0.65) {};
    \node (arout1) at (-1.05,0.75) {};
    \node (arout2) at (-1.26,0.9) {};
    \draw[-] (U) -- (n1);
    \draw[-] (U) -- (v) -- (n2);
    \draw[-] (U) -- (u) -- (n3);
    \draw[-] (U) -- (n4);
    \draw[->] (arin1) -- (arin2);
    \draw[->] (arout2) -- (arout1);
    \end{tikzpicture}
\end{equation}
Here, $R[w]$ is the adjoint, 3-dimensional representation of $w\in \mathrm{SU}(2)$ and 
\begin{equation}
    D = \begin{pmatrix}
    \sin 2J & 0 & 0\cr 
    0 & \sin 2J & 0\cr
    0 & 0 & 1
    \end{pmatrix}.
\end{equation}
Therefore, the moment of the correlation function for a fixed mikado arrangement reads
\begin{eqnarray}
C^{(k)}_{ab}(x_{\rm i},t_{\rm i};x_{\rm f},t_{\rm f}) &=& 
\sum_{n}
\delta(x_{\rm i}-x_n(t_{\rm i}))\delta(x_{\rm f}-x_{n}(t_{\rm f})) \label{eq:cff2} \\
&&\; \times\bra{a}\mathcal M[u_m,v_m]\cdots
\mathcal M[u_2,v_2]\mathcal M[u_1,v_1]\ket{b}^k,
\nonumber
\end{eqnarray}
where $\bra{a},\ket{b}$ are 3-dimensional vectors representing operators $a,b$ in the Pauli basis, while $u_j$, and $v_j$, $j=1,\ldots,m$ are SU(2) 1-qubit gates before, and after, the interaction vertices (in parametrization (\ref{du_unit})) along all $m=m(n)$ intersections of the line/wire $n$ with all the other lines between times $t_{\rm i}$ and $t_{\rm f}$.

Finally, we write the average moment of spatio-temporal correlator as an expectation value over random mikado arrangements and Haar random gates $u_j,v_j$, while the interaction parameter $J$ is kept fixed as the key parameter (coupling constant) of the model:
\begin{eqnarray}
\mathcal C^{(k)}_{ab}(x,t) &=& \mathbb E\left(
C^{(k)}_{ab}(0,0;x,t)\right)
= \frac{1}{t}p\left(\frac{x}{t}\right)\sum_{m=0}^\infty 
\mathbb P(m|x,t) \mathcal C^{(k)}_{ab}(m).
\label{eq:C}
\end{eqnarray}
In random mikado arrangements we assume translational invariance in $x$ and $t$, hence $N=\infty$, so we can fix 
$x_{\rm i}=0$, $t_{\rm i}=0$.
$p(v)$ denotes the probability density that a randomly chosen line ($n$) has the velocity/slope $v_n=v$, while 
$\mathbb P(m|x,t)$ denotes the probability that a line section from $(0,0)$ to $(x,t)$ has exactly $m$ crossings with other lines, while 
$C^{(k)}_{ab}(m)$ denotes an average over Haar random $u_j,v_j$:
\begin{equation}
\mathcal C^{(k)}_{ab}(m) =
 \mathbb E \left(\bra{a}\mathcal M[u_m,v_m]\cdots
\mathcal M[u_2,v_2]\mathcal M[u_1,v_1]\ket{b}^k\right).
\label{Ck}
\end{equation}
We can evaluate the expectation (\ref{Ck}) by using $k$ replicas
and explicit Haar measure for integration over Euler angle parametrization of the rotation matrix $R[u]\equiv R(\vec{\theta}) \equiv R(\theta_1,\theta_2,\theta_3)$,
\begin{equation}
    \int d\mu(\Vec{\theta}) = \frac{1}{8\pi^2} 
    \int_0^{2\pi} d\theta_1
    \int_0^\pi d\theta_2 
    \int_0^{2\pi} d\theta_3 \sin(\theta_2).
\end{equation}
Specifically,
\begin{eqnarray}
\mathcal C^{(k)}_{ab}(m) &=&
\bra{a}^{\otimes k}
\int \prod_{1\le j\le m}^{\leftarrow} d\mu(\vec{\theta}_j)
d\mu(\vec{\theta}'_j) 
R(\vec{\theta}_j)^{\otimes k}D^{\otimes k} 
R(\vec{\theta}'_j)^{\otimes k}
\ket{b}^{\otimes k} \nonumber \\
&=& \bra{a}^{\otimes k} \mathcal T^m 
\ket{b}^{\otimes k},
\end{eqnarray}
where $\mathcal T$ is the $3^k \times 3^k$
transfer matrix
\begin{equation}
    \mathcal T = \mathcal P 
     D^{\otimes k} \mathcal P,\quad
     \mathcal P:= 3^{-k/2}\int d\mu(\vec{\theta})
     R(\vec{\theta})^{\otimes k}.
\end{equation}
While the projector $\mathcal P$ can be in general expressed in terms of the Weingarten functions~\cite{Weingarten,Collins_2003}, the results for the first and the second moment can be obtained fully explicitly.

Let us assume that $a,b$ is an arbitrary pair of traceless and Hilbert-Schmidt normalised observables, $\tr a=\tr b =0$, $\tr a^2=\tr b^2 =1$. 

For $k=1$, we find $\mathcal P = 0$,
and hence the average (mean) correlator vanishes for any $m>0$,
\begin{equation}
    \mathcal C^{(1)}_{ab}(m) = \delta_{m,0}.
\end{equation}
For $k=2$, we find that
\begin{equation}
\mathcal P = \frac{1}{3}
    \begin{pmatrix}
1 & 0 & 0 & 0 & 1  & 0 & 0 & 0 & 1 \\
    0 & 0 & 0 & 0 & 0 & 0 & 0 & 0 & 0 \\
    0 & 0 & 0 & 0 & 0 & 0 & 0 & 0 & 0 \\
    0 & 0 & 0 & 0 & 0 & 0 & 0 & 0 & 0 \\
    1 & 0 & 0 & 0 & 1  & 0 & 0 & 0 & 1 \\
    0 & 0 & 0 & 0 & 0 & 0 & 0 & 0 & 0 \\
    0 & 0 & 0 & 0 & 0 & 0 & 0 & 0 & 0 \\
    0 & 0 & 0 & 0 & 0 & 0 & 0 & 0 & 0 \\
   1 & 0 & 0 & 0 & 1 & 0 & 0 & 0 & 1
    \end{pmatrix},
\end{equation}
and the transfer matrix has rank one
\begin{equation}
    \mathcal T = \lambda \ket{\Psi}\bra{\Psi},\quad
    \lambda = \frac{1}{3}(2-\cos 4J),\quad
    \ket{\Psi} = \frac{1}{\sqrt{3}}\sum_{\alpha=1}^3 \ket{\alpha}\otimes\ket{\alpha}.
\end{equation}
Hence the second moment of the correlator reads
\begin{equation}
\label{eq:cor_second_mom}
\mathcal C^{(2)}_{ab}(m) =
\delta_{m,0}+(1-\delta_{m,0})
\frac{1}{3}\lambda^m = 
\delta_{m,0}+(1-\delta_{m,0)})
\frac{1}{3^{m+1}}( 2- \cos(4J))^m\,,
\end{equation}
while the variance is given by:
\begin{equation}
\label{eq:cor_variance}
    \mathcal C^{[2]}_{ab}(m) = \mathcal C^{(2)}_{ab}(m) - \left(\mathcal C^{(1)}_{ab}(m)\right)^2 = (1-\delta_{m,0)})
\frac{1}{3^{m+1}}( 2- \cos(4J))^m\,.
\end{equation}

\section{Two settings of random mikado networks}
\label{sec:settings}
In this section, we will introduce the two settings of random networks that we studied.
In the first setting, studied in \cref{Gaussian_wires}, we assume to have a gas of $N\to\infty$ identical free non-interacting classical point particles in thermal equilibrium, at fixed temperature and density. We then assume that particles carry internal quantum degrees of freedom - qubits, which interact via random dual unitary gate whenever two particles meet (and their trajectories cross).

In the second setting, studied in \cref{subsec:RAND_BACK}, we consider a rotationally and translationally invariant random arrangement of wires, such that intersections with any, arbitrarily oriented fixed line have a constant density.

Distinct outputs of these settings are conditional number-of-crossed-wires probabilities $\mathbb P(m|x,t)$ which is the essential input into the formula (\ref{eq:C}).

\subsection{Thermal distribution of wires}
\label{Gaussian_wires}

Here we consider a gas of $N\to\infty$ non-interacting particles with coordinates
$x_n(t) = x_n + v_n t$. We assume the gas to be in thermal equilibrium, with mean interparticle spacing $a=1$ (or density one), setting the length scale, and inverse temperature $\beta=1$, setting the time scale. This means that coordinates of particles $x_n(t)$ form a Poisson point process on $\mathbb R$  with mean spacing $a=1$ for all times $t$, and that velocities $v_n$ are i.i.d. normal Gaussian variables with zero mean and unit variance
\begin{equation}
    p(v) = \frac{1}{\sqrt{2\pi}}e^{-v^2/2}\,.
    \label{eq:gauss}
\end{equation}
Given a random line $x_n(t)$, $t\in[0,T]$, we shall now write the 
probability distribution $\mathbb P(m|X,T)$, where $X=v_n T$, of the number $m$ of other lines $x_{n'}(t)$ crossing the line $n$ during the time $t\in[0,T]$. As the wires are statistically independent, the distribution is clearly the Poissonian
\begin{equation}
    \mathbb P(m|X,T)=\frac{[\bar{m}(X,T)]^m e^{-\bar{m}(X,T)}}{m!}.
    \label{Poissonian}
\end{equation}
We shall now derive the average number of crossings $\bar{m}(X,T)$. 

We first write the probability of crossing of a pair of randomly chosen wires, indicated as $n=0$ and $n=1$, for $t\in[0,T]$. For convenience, we choose the reference initial coordinate at the origin $x_0=0$, and the second one at $x_1=x$, with slopes/velocities, $v_0,v_1$, respectively.
The condition for crossing in the targeted time interval then 
reads $0 < x/(v_0-v_1)< T$, which gives the probability, after integrating over the normally distributed velocity $v_1$
\begin{equation}
\begin{split}
    \mathbb P(v_0,x,T) =& \frac{1}{\sqrt{2\pi}} \left( \int_{-\infty}^{x/T - v_0} e^{-\frac{1}{2}v_1^2} + \int^{\infty}_{x/T - v_0} e^{-\frac{1}{2}v_1^2} \right) dv_1 \\
 &= \frac{1}{2} + \frac{\mathrm{sgn}(x)}{2} \mathrm{erf} \left( \frac{1}{\sqrt{2}} \left( v_0 - \frac{x}{T}\right)\right).
\end{split}
\end{equation}
As the initial coordinates $x$ of the wires are distributed uniformly randomly on $\mathbb R$ with density one, the average number of crossings then reads (noting $v_0=X/T$):
\begin{equation}
\label{eq:av_num_thermal}
    \bar{m}(X,T) = \int_{-\infty}^\infty\!dx\,\mathbb P\left(\frac{X}{T},x,T\right) \\
    =  \sqrt{\frac{2}{\pi}}\,T \exp\left(-\frac{X^2}{2T^2}\right) + 
    X\,\mathrm{erf}\left( \frac{X}{\sqrt{2}T}\right).
\end{equation}

\subsection{Isotropic distribution of wires}
\label{subsec:RAND_BACK}
In the second setting, we assume that the wire distribution is isotropic and translationally invariant, with density one.
This immediately implies that the slope (velocity) distribution is given by a distribution of a tangent of a uniformly distributed angle
\begin{equation}
    p(v) = \frac{1}{\pi(1+v^2)}\,,
    \label{eq:tan}
\end{equation}
while the distribution of the number
of crossings is again Poissonian (\ref{Poissonian}), with the average that is given by the length of line-section (wire)
\begin{equation}
    \bar{m}(X,T) = \sqrt{X^2+T^2}\,.
    \label{eq:pitagora}
\end{equation}

\begin{figure}
  \centering
  \begin{subfigure}{.9\textwidth}
        \centering
        \includegraphics[width=1\linewidth]{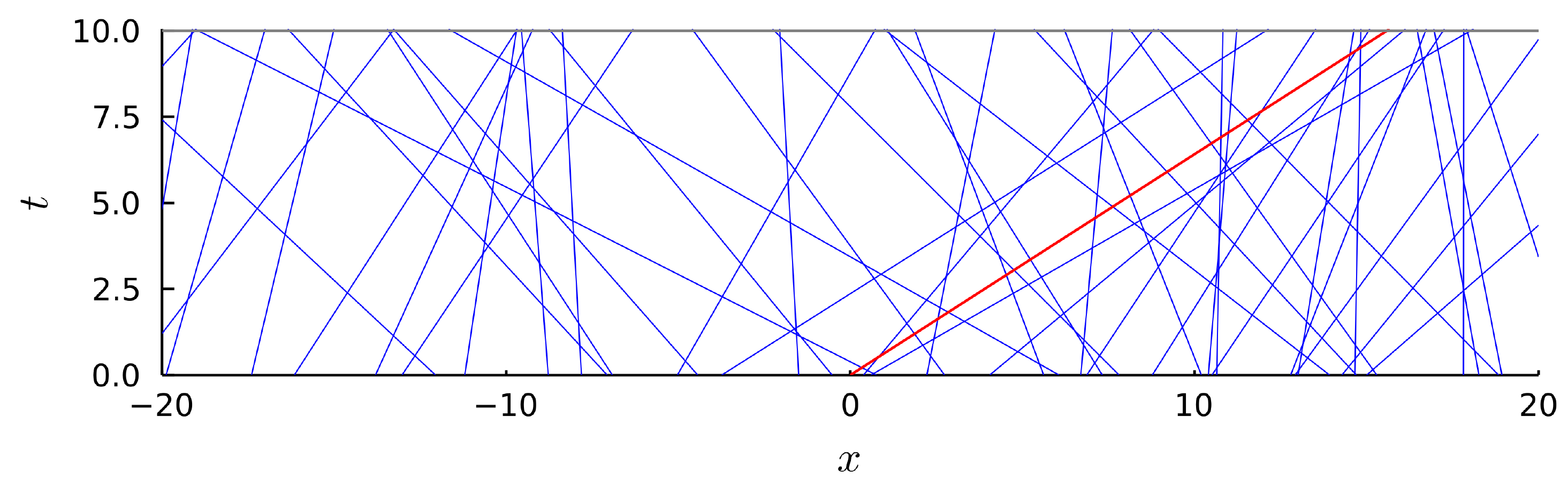}
        \caption{Example of a realization
        of a mikado circuit for the case of thermal distribution of slopes.} 
        \label{fig:G_wire}
  \end{subfigure}\\
  
  \begin{subfigure}{.9\textwidth}
        \centering
        \includegraphics[width=1\linewidth]{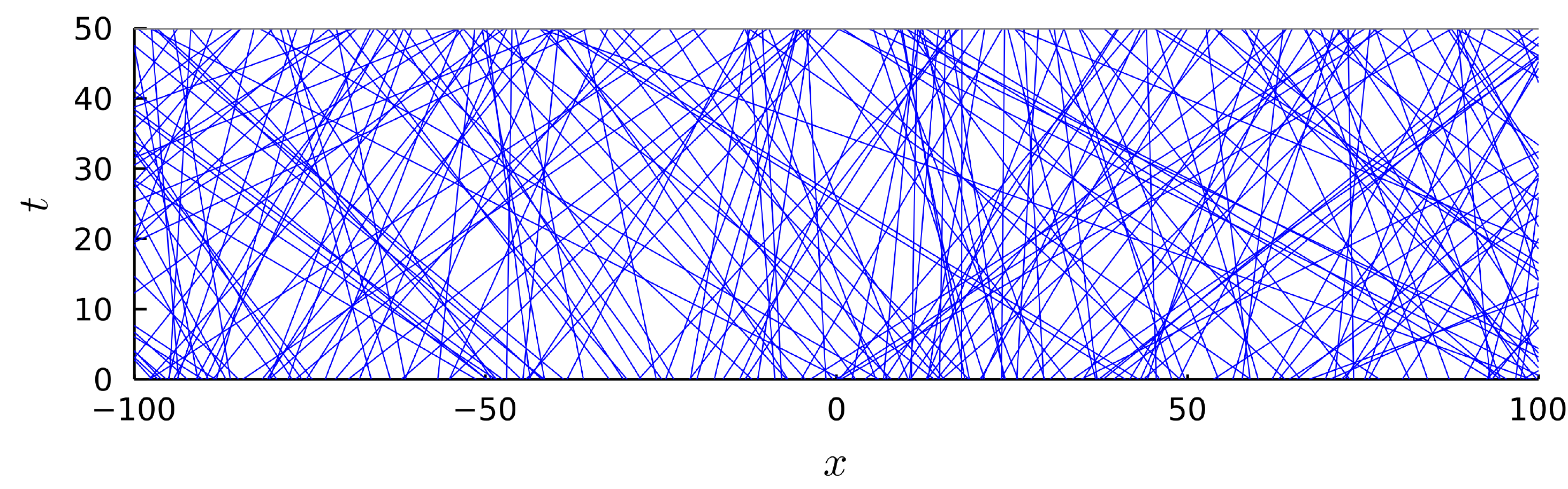}
        \caption{Zoom out of the thermal case.}
        \label{fig:thermal_unzoom}
  \end{subfigure}\\
  
  \begin{subfigure}{.9\textwidth}
        \centering
        \includegraphics[width=1\linewidth]{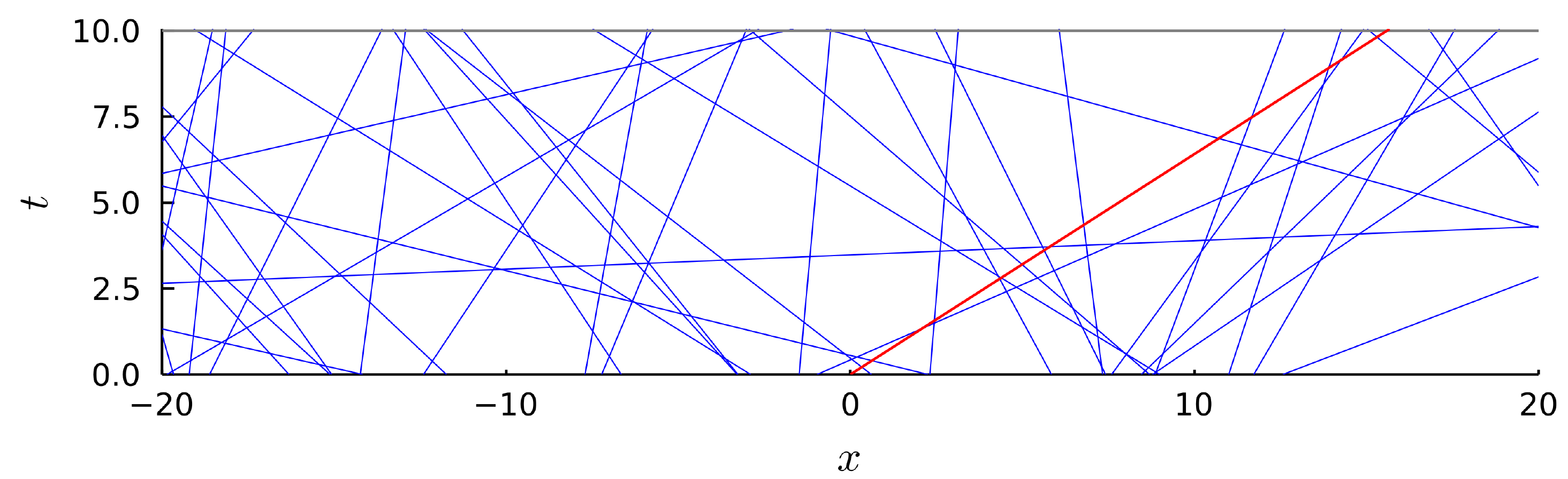}
        \caption{Example of a realization
        of a mikado circuit for the case of isotropic distribution of slopes.}
        \label{fig:RB_Z}
  \end{subfigure}\\
  
  \begin{subfigure}{.9\textwidth}
        \centering
        \includegraphics[width=1\linewidth]{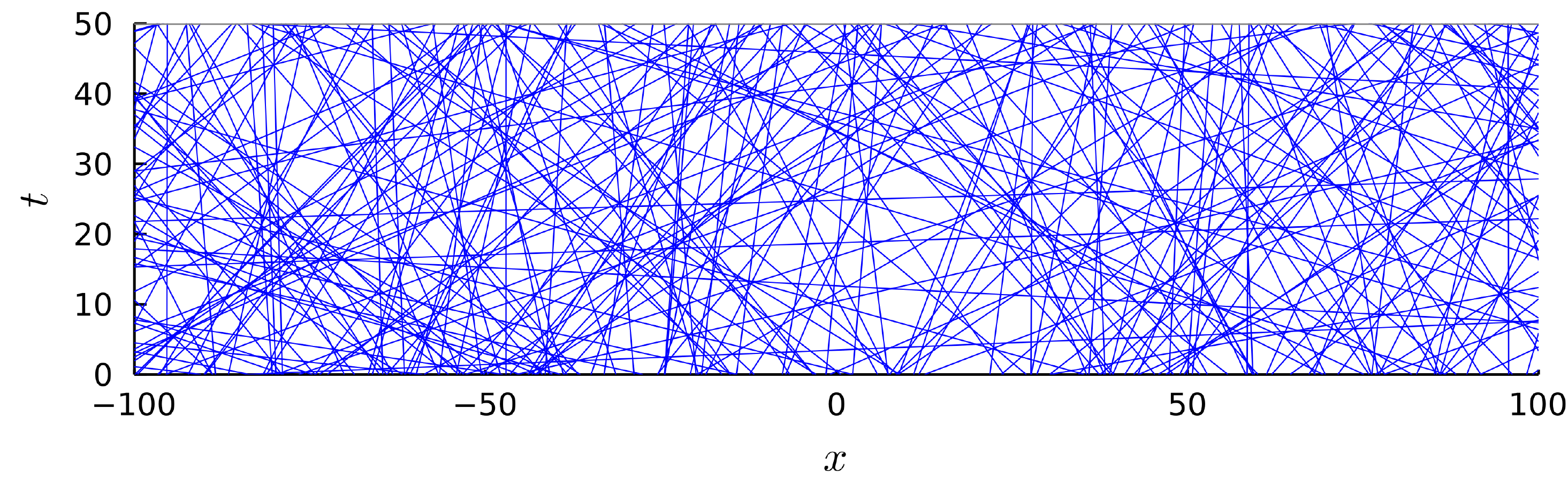}
        \caption{Zoom out of the isotropic case.}
        \label{fig:iso_unzoom}
  \end{subfigure}
 \caption{Example of the two types of random mikado geometries studied.}
\end{figure}


\section{The variance of spatio-temporal correlation function}

\begin{figure}
  \centering
  \begin{subfigure}{.9\textwidth}
        \centering
        \includegraphics[width=1\linewidth]{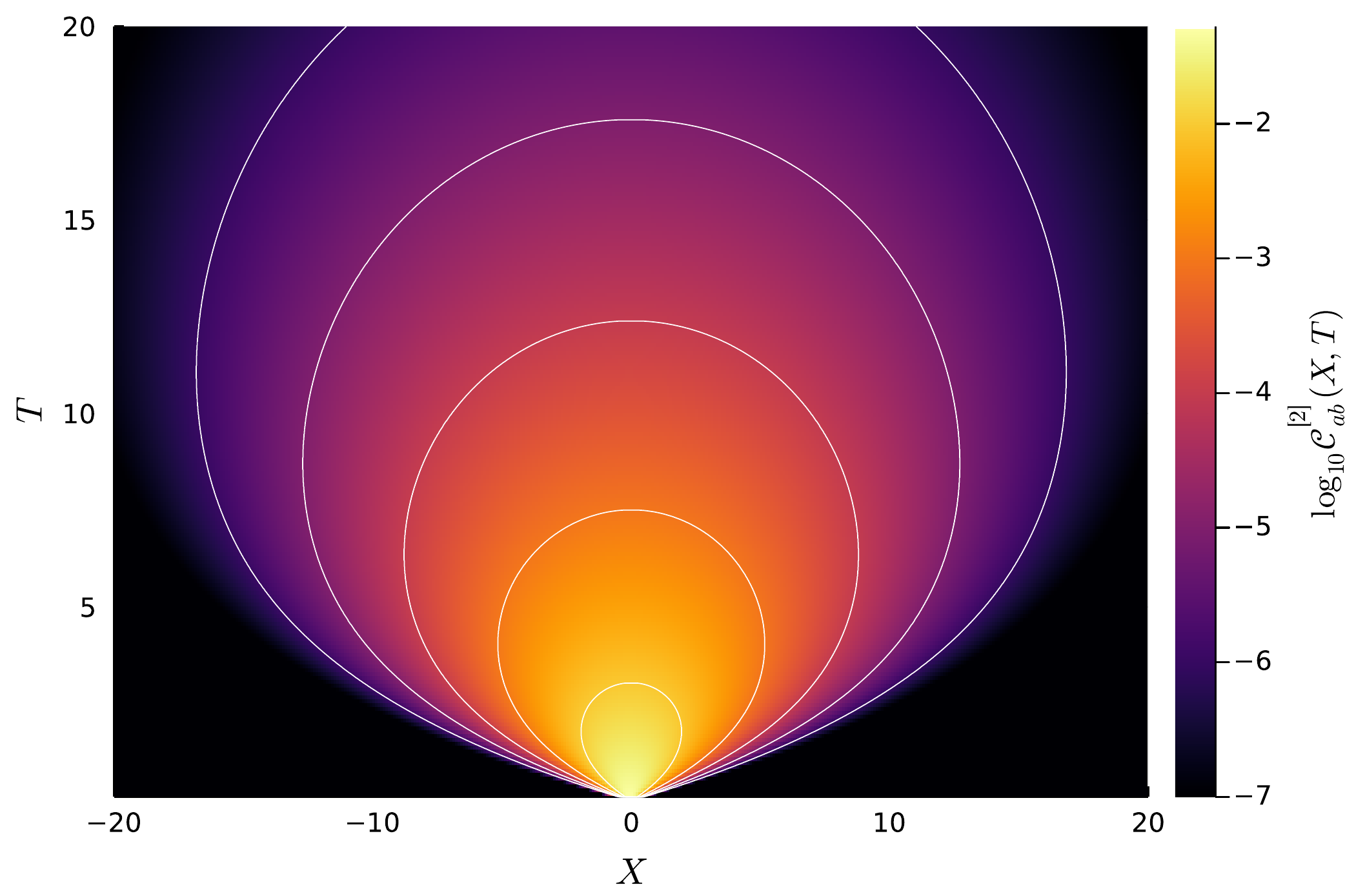}
        \caption{Heatmap for the thermal case Eq.~(\ref{eq:thermalC}).} 
        \label{fig:corr_full_thermal}
  \end{subfigure}\\
  
  \begin{subfigure}{.9\textwidth}
        \centering
        \includegraphics[width=1\linewidth]{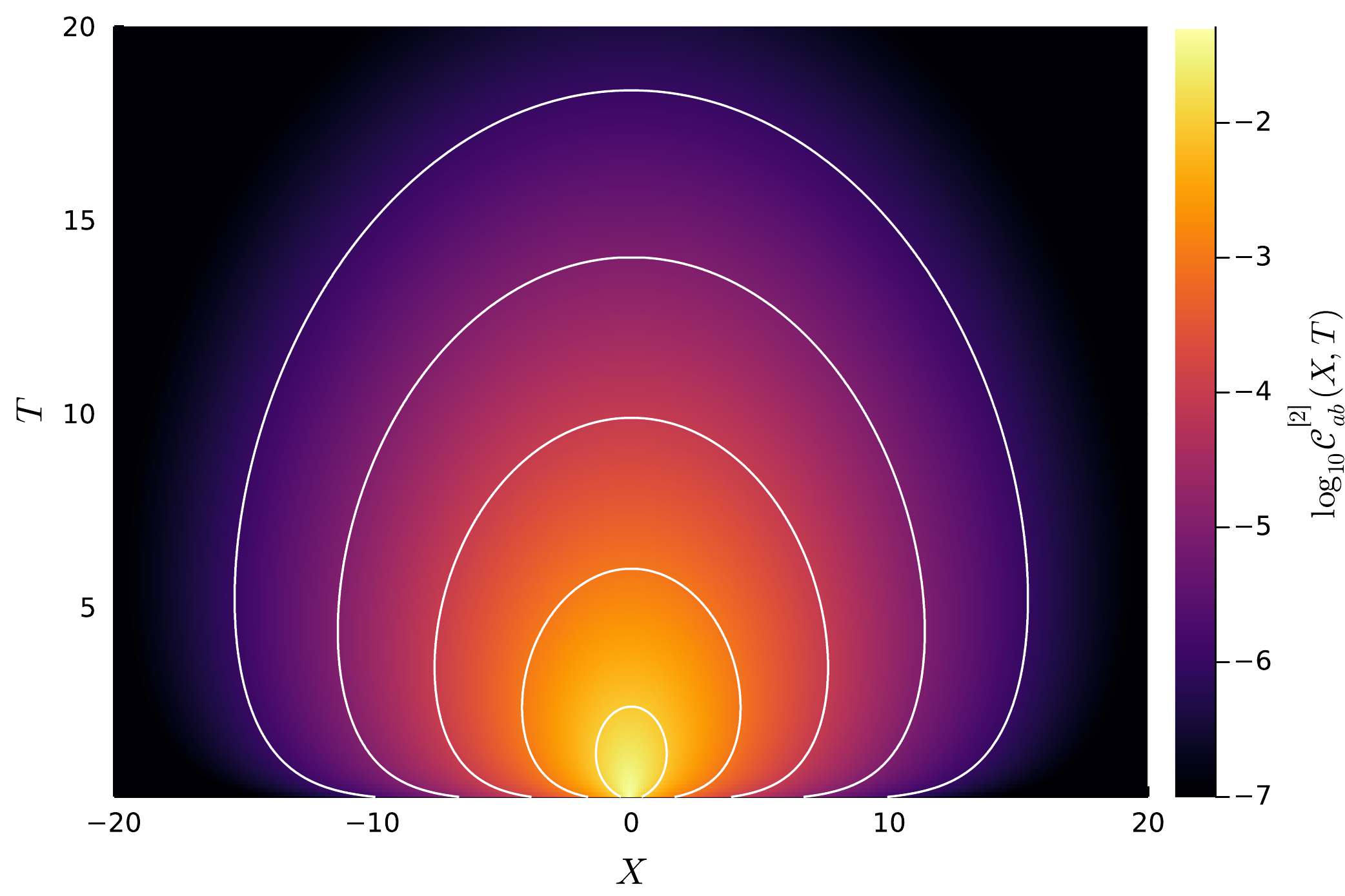}
        \caption{Heatmap for the isotropic case Eq.~(\ref{eq:IsoC})}
        \label{fig:corr_full_iso}
  \end{subfigure}
  \caption{Heatmaps of the fluctuation of the correlation functions with $J=\pi/4-1/2$ in log-scale, Eq.~(\ref{eq:final_corr}). The value of the variance of the correlation function are constant along the iso-contours drawn in white. We note that we took a cutoff for the minimum values shown at $\log_{10} \mathcal{C}^{[2]}_{ab}(X,T) = -7$.}
  \label{fig:Heatmaps_variance}
  \end{figure}

Finally, we collect the pieces (\ref{eq:C},\ref{eq:cor_variance},\ref{Poissonian}) together, and write the final result for the variance of local 2-point correlation function in mikado dual unitary circuits:
\begin{equation}
\label{eq:final_corr}
\begin{split}
    \mathcal{C}^{[2]}_{ab}(X,T) &= \frac{1}{T}\, p\!\left(\frac{X}{T}\right)\sum_{m=0}^\infty (1-\delta_{m,0)})\frac{1}{3^{m+1}}\frac{[\bar{m}(X,T)]^m e^{-\bar{m}(X,T)}}{m!} (2-\cos 4J)^m \\
    &= \frac{1}{3T}\, p\!\left(\frac{X}{T}\right)\sum_{m=1}^\infty \frac{[\bar{m}(X,T)]^m e^{-\bar{m}(X,T)}}{m!} \left(1-\frac{2}{3}\cos^2(2J)\right)^m \\
    &=\frac{1}{3T}\, p\!\left(\frac{X}{T}\right)\left\{  \exp \left( -\frac{2}{3}\cos^2(2J)\bar{m}(X,T)\right) - \exp\left( -\bar{m}(X,T) \right)\right\}\,.
\end{split}
\end{equation}
For the two different distributions of random mikados that we have studied in \cref{sec:settings} we can write the result even more explicitly:

i) For the thermal case, we use (\ref{eq:gauss},\ref{eq:av_num_thermal}), and find
\begin{equation}
    \begin{split}
        \mathcal{C}^{[2]}_{ab}(X,T) &=
        \frac{1}{3\sqrt{2\pi}T}
        \exp\! \left(-
        \frac{X^2}{2T^2} \right)
        \left\{
        \exp \! \left[ -\frac{2\! \cos^2 \!2J}{3} \left\{ \sqrt{\frac{2}{\pi}} T e^{-\frac{X^2}{2T^2}}
         + X \mathrm{erf}\left( \frac{X}{\sqrt{2}T} \right) \right\} \right] \right. \\
         & \left.  - \exp \left[- 
         \sqrt{\frac{2}{\pi}} T e^{-\frac{X^2}{2T^2}}
         - X \mathrm{erf} \left( \frac{X}{\sqrt{2}T} \right)
         \right]
        \right\}
    \end{split}
\label{eq:thermalC}
\end{equation}

ii) For the isotropic case, the result is even simpler via
(\ref{eq:tan},\ref{eq:pitagora})
\begin{equation}
    \mathcal{C}^{[2]}_{ab}(X,T) = 
    \frac{T}{3\pi(X^2+T^2)}
    \left\{ \exp\left(
    -\frac{2 \cos^2 2J}{3}\sqrt{X^2+T^2}\right) - \exp\left( - \sqrt{X^2+T^2} \right)\right \}\,.
\label{eq:IsoC}
\end{equation}
For illustration, we show in \cref{fig:Heatmaps_variance} the variance of the correlator for an exemplar value of the interaction parameter $J=\pi/4 - 1/2$. Specifically, \cref{fig:corr_full_thermal}
depicts the correlator in the thermal case (\ref{eq:thermalC}), and
\cref{fig:corr_full_iso} in the isotropic case (\ref{eq:IsoC}).

\section{Discussion}

We proposed and explored random dual unitary circuits in random geometries. We studied two settings of the so-called mikado random geometries, one with a \emph{preferred} space-time direction and the other with a \emph{random} isotropic space-time. 
The first one can be physically motivated in terms of a thermalised ideal gas of identical classical point particles carrying quantum qubit degrees of freedom, which experience dual unitary scattering. The second one is a curious example of a euclidean space-time with equivalent unitary-quantum dynamics in a continuum set of directionalities. In both cases we have shown that, while the average local correlator vanishes due to randomness of the local gates, the variance of the correlator can be computed explicitly in terms of a simple rank-one transfer matrix. Computation of higher moments of the correlator involves a nontrivial Weingarten calculus and shall be left for future work.

Unitary dynamics in a continuum set of space-time directions found here should be contrasted with a pair of unitary directions in
regular brickwork dual unitary circuits~\cite{Bertini_et_al_PRL_2019} or a triple of unitary space-time directions in the so-called tri-unitary circuits~\cite{Khemani}.

While this work addressed the simplest problem
of computation of local correlations in mikado dual unitary circuits, there are several immediate pending questions for future explorations.
For example, it would be interesting to compute entanglement dynamics, both for states and operators, 
study perturbed non-dual-unitary defects, or even attempt to extend the concepts of dual unitaries on random geometries to higher dimensions (in analogy to recent study~\cite{mendl}). 

Last, but not least, one may wonder if an extension of random circuit geometry to arbitrary (non-mikado) graphs with degree 4  (i.e.  {\em random 4-regular graphs}) would be feasible. Perhaps one can find a meaningful ensemble of planar random 4-regular graphs for this purpose. Clearly, in such cases, space-time paths which carry quantum information are no-longer straight lines, and interesting physical effects may occur when such paths can form self-intersections, or loops.


\section*{Acknowledgements}
We thank Bruno Bertini, Felix Fritzsch and Pavel Kos for fruitful discussions. This research has received funding from the European Union's Horizon 2020 research and innovation programme under the Marie Skłodowska-Curie grant agreement number 955479, and under ERC Advanced Grant 694544--OMNES, as well as from Slovenian Research agency (ARRS) under programme P1-0402.

\printbibliography

@article{Bertini_et_al_PRL_2019,
  title = {Exact Correlation Functions for Dual-Unitary Lattice Models in $1+1$ Dimensions},
  author = {Bertini, Bruno and Kos, Pavel and Prosen, Tomaž},
  journal = {Phys. Rev. Lett.},
  volume = {123},
  issue = {21},
  pages = {210601},
  numpages = {6},
  year = {2019},
  month = {Nov},
  publisher = {American Physical Society},
  doi = {10.1103/PhysRevLett.123.210601},
  url = {https://link.aps.org/doi/10.1103/PhysRevLett.123.210601}
}

@article{BKP18,
  title = {Exact Spectral Form Factor in a Minimal Model of Many-Body Quantum Chaos},
  author = {Bertini, Bruno and Kos, Pavel and Prosen, Tomaz},
  journal = {Phys. Rev. Lett.},
  volume = {121},
  issue = {26},
  pages = {264101},
  numpages = {6},
  year = {2018},
  month = {Dec},
  publisher = {American Physical Society},
  doi = {10.1103/PhysRevLett.121.264101},
  url = {https://link.aps.org/doi/10.1103/PhysRevLett.121.264101}
}

@article{CMP,
  title={Random matrix spectral form factor of dual-unitary quantum circuits},
  author={Bertini, Bruno and Kos, Pavel and Prosen, Toma{\v{z}}},
  journal={Communications in Mathematical Physics},
  volume={387},
  number={1},
  pages={597--620},
  year={2021},
  publisher={Springer}
}

@article{Piroli,
  title = {Exact dynamics in dual-unitary quantum circuits},
  author = {Piroli, Lorenzo and Bertini, Bruno and Cirac, J. Ignacio and Prosen, Tomaz},
  journal = {Phys. Rev. B},
  volume = {101},
  issue = {9},
  pages = {094304},
  numpages = {16},
  year = {2020},
  month = {Mar},
  publisher = {American Physical Society},
  doi = {10.1103/PhysRevB.101.094304},
  url = {https://link.aps.org/doi/10.1103/PhysRevB.101.094304}
}

@article{Khemani,
  title = {Triunitary quantum circuits},
  author = {Jonay, Cheryne and Khemani, Vedika and Ippoliti, Matteo},
  journal = {Phys. Rev. Research},
  volume = {3},
  issue = {4},
  pages = {043046},
  numpages = {14},
  year = {2021},
  month = {Oct},
  publisher = {American Physical Society},
  doi = {10.1103/PhysRevResearch.3.043046},
  url = {https://link.aps.org/doi/10.1103/PhysRevResearch.3.043046}
}

@article{Kos_et_al_PRX_2021,
  title = {Correlations in Perturbed Dual-Unitary Circuits: Efficient Path-Integral Formula},
  author = {Kos, Pavel and Bertini, Bruno and Prosen, Tomaž},
  journal = {Phys. Rev. X},
  volume = {11},
  issue = {1},
  pages = {011022},
  numpages = {30},
  year = {2021},
  month = {Feb},
  publisher = {American Physical Society},
  doi = {10.1103/PhysRevX.11.011022},
  url = {https://link.aps.org/doi/10.1103/PhysRevX.11.011022}
}

@Article{Bertini_et_al_SCIPOST_2020,
	title={{Operator Entanglement in Local Quantum Circuits I: Chaotic Dual-Unitary  Circuits}},
	author={Bruno Bertini and Pavel Kos and Tomaž Prosen},
	journal={SciPost Phys.},
	volume={8},
	issue={4},
	pages={67},
	year={2020},
	publisher={SciPost},
	doi={10.21468/SciPostPhys.8.4.067},
	url={https://scipost.org/10.21468/SciPostPhys.8.4.067},
}

@article{MCBanuls_et_al_2009,
  title = {Matrix Product States for Dynamical Simulation of Infinite Chains},
  author = {Ba\~nuls, M. C. and Hastings, M. B. and Verstraete, F. and Cirac, J. I.},
  journal = {Phys. Rev. Lett.},
  volume = {102},
  issue = {24},
  pages = {240603},
  numpages = {4},
  year = {2009},
  month = {Jun},
  publisher = {American Physical Society},
  doi = {10.1103/PhysRevLett.102.240603},
  url = {https://link.aps.org/doi/10.1103/PhysRevLett.102.240603}
}

@article{Arul1,
  title = {Creating Ensembles of Dual Unitary and Maximally Entangling Quantum Evolutions},
  author = {Rather, Suhail Ahmad and Aravinda, S. and Lakshminarayan, Arul},
  journal = {Phys. Rev. Lett.},
  volume = {125},
  issue = {7},
  pages = {070501},
  numpages = {6},
  year = {2020},
  month = {Aug},
  publisher = {American Physical Society},
  doi = {10.1103/PhysRevLett.125.070501},
  url = {https://link.aps.org/doi/10.1103/PhysRevLett.125.070501}
}

@article{Arul2,
  title = {From dual-unitary to quantum Bernoulli circuits: Role of the entangling power in constructing a quantum ergodic hierarchy},
  author = {Aravinda, S. and Rather, Suhail Ahmad and Lakshminarayan, Arul},
  journal = {Phys. Rev. Research},
  volume = {3},
  issue = {4},
  pages = {043034},
  numpages = {30},
  year = {2021},
  month = {Oct},
  publisher = {American Physical Society},
  doi = {10.1103/PhysRevResearch.3.043034},
  url = {https://link.aps.org/doi/10.1103/PhysRevResearch.3.043034}
}

@article{Claeys1,
  title = {Maximum velocity quantum circuits},
  author = {Claeys, Pieter W. and Lamacraft, Austen},
  journal = {Phys. Rev. Research},
  volume = {2},
  issue = {3},
  pages = {033032},
  numpages = {20},
  year = {2020},
  month = {Jul},
  publisher = {American Physical Society},
  doi = {10.1103/PhysRevResearch.2.033032},
  url = {https://link.aps.org/doi/10.1103/PhysRevResearch.2.033032}
}

@article{Claeys2,
  title = {Ergodic and Nonergodic Dual-Unitary Quantum Circuits with Arbitrary Local Hilbert Space Dimension},
  author = {Claeys, Pieter W. and Lamacraft, Austen},
  journal = {Phys. Rev. Lett.},
  volume = {126},
  issue = {10},
  pages = {100603},
  numpages = {6},
  year = {2021},
  month = {Mar},
  publisher = {American Physical Society},
  doi = {10.1103/PhysRevLett.126.100603},
  url = {https://link.aps.org/doi/10.1103/PhysRevLett.126.100603}
}

@article{Prosen21,
  title={Many-body quantum chaos and dual-unitarity round-a-face},
  author={Prosen, Toma{\v{z}}},
  journal={Chaos: An Interdisciplinary Journal of Nonlinear Science},
  volume={31},
  number={9},
  pages={093101},
  year={2021},
  publisher={AIP Publishing LLC}
}

@article{Google2019,
  doi = {10.1038/s41586-019-1666-5},
  url = {https://doi.org/10.1038/s41586-019-1666-5},
  year = {2019},
  month = oct,
  publisher = {Springer Science and Business Media {LLC}},
  volume = {574},
  number = {7779},
  pages = {505--510},
  author = {Frank Arute and Kunal Arya and Ryan Babbush and Dave Bacon and Joseph C. Bardin and Rami Barends and Rupak Biswas and Sergio Boixo and Fernando G. S. L. Brandao and David A. Buell and Brian Burkett and Yu Chen and Zijun Chen and Ben Chiaro and Roberto Collins and William Courtney and Andrew Dunsworth and Edward Farhi and Brooks Foxen and Austin Fowler and Craig Gidney and Marissa Giustina and Rob Graff and Keith Guerin and Steve Habegger and Matthew P. Harrigan and Michael J. Hartmann and Alan Ho and Markus Hoffmann and Trent Huang and Travis S. Humble and Sergei V. Isakov and Evan Jeffrey and Zhang Jiang and Dvir Kafri and Kostyantyn Kechedzhi and Julian Kelly and Paul V. Klimov and Sergey Knysh and Alexander Korotkov and Fedor Kostritsa and David Landhuis and Mike Lindmark and Erik Lucero and Dmitry Lyakh and Salvatore Mandr{\`{a}} and Jarrod R. McClean and Matthew McEwen and Anthony Megrant and Xiao Mi and Kristel Michielsen and Masoud Mohseni and Josh Mutus and Ofer Naaman and Matthew Neeley and Charles Neill and Murphy Yuezhen Niu and Eric Ostby and Andre Petukhov and John C. Platt and Chris Quintana and Eleanor G. Rieffel and Pedram Roushan and Nicholas C. Rubin and Daniel Sank and Kevin J. Satzinger and Vadim Smelyanskiy and Kevin J. Sung and Matthew D. Trevithick and Amit Vainsencher and Benjamin Villalonga and Theodore White and Z. Jamie Yao and Ping Yeh and Adam Zalcman and Hartmut Neven and John M. Martinis},
  title = {Quantum supremacy using a programmable superconducting processor},
  journal = {Nature}
}

@article{Collins_2003,
    author = {Collins, Benoît},
    title = "{Moments and cumulants of polynomial random variables on unitarygroups, the Itzykson-Zuber integral, and free probability}",
    journal = {International Mathematics Research Notices},
    volume = {2003},
    number = {17},
    pages = {953-982},
    year = {2003},
    month = {01},
    issn = {1073-7928},
    doi = {10.1155/S107379280320917X},
    url = {https://doi.org/10.1155/S107379280320917X},
    eprint = {https://academic.oup.com/imrn/article-pdf/2003/17/953/1881428/2003-17-953.pdf},
}

@misc{mendl,
  doi = {10.48550/ARXIV.2206.01499},
  
  url = {https://arxiv.org/abs/2206.01499},
  
  author = {Milbradt, Richard and Scheller, Lisa and Aßmus, Christopher and Mendl, Christian B.},
  
  title = {Ternary unitary quantum lattice models and circuits in $2 + 1$ dimensions},
  
  publisher = {arXiv},
  
  year = {2022},
  
  copyright = {arXiv.org perpetual, non-exclusive license}
}

@article{Weingarten,
author = {Weingarten,Don },
title = {Asymptotic behavior of group integrals in the limit of infinite rank},
journal = {Journal of Mathematical Physics},
volume = {19},
number = {5},
pages = {999-1001},
year = {1978},
doi = {10.1063/1.523807},

URL = { 
        https://doi.org/10.1063/1.523807
    
},
eprint = { 
        https://doi.org/10.1063/1.523807
    
}

}

@article{Marton,
  title = {Hybrid Semiclassical Theory of Quantum Quenches in One-Dimensional Systems},
  author = {Moca, C. P. and Kormos, M\'arton and Zar\'and, Gergely},
  journal = {Phys. Rev. Lett.},
  volume = {119},
  issue = {10},
  pages = {100603},
  numpages = {5},
  year = {2017},
  month = {Sep},
  publisher = {American Physical Society},
  doi = {10.1103/PhysRevLett.119.100603},
  url = {https://link.aps.org/doi/10.1103/PhysRevLett.119.100603}
}

\end{document}